\documentclass[letter]{aa} % for the letters 
\usepackage{amsmath}
\usepackage{graphicx}
\usepackage{graphics}
\usepackage{txfonts}
\usepackage{natbib}
\usepackage{xcolor}
\bibpunct{(}{)}{;}{a}{}{,}

\def\teff{\ifmmode T_{\rm eff} \else $T_{\mathrm{eff}}$\fi}

\def\ltsima{$\buildrel<\over\sim$}
\def\lsim{\lower.5ex\hbox{\ltsima}}

\newcommand{\ha}{\ifmmode {\rm H}\alpha \else H$\alpha$\fi}
\newcommand{\hb}{\ifmmode {\rm H}\beta \else H$\beta$\fi}
\newcommand{\lya}{\ifmmode {\rm Ly}\alpha \else Ly$\alpha$\fi}

\newcommand{\heii}{He~{\sc ii}}
\newcommand{\Heiiuv}{He~{\sc ii} $\lambda$1640}
\newcommand{\Heiiopt}{He~{\sc ii} $\lambda$4686}

\newcommand{\ebv}{\ifmmode E_{\rm B-V} \else $E_{\rm B-V}$\fi}
\newcommand{\av}{\ifmmode A_{\rm V} \else $A_{\rm V}$\fi}
\def\kms{km s$^{-1}$}

\def\msun{\ifmmode M_{\odot} \else M$_{\odot}$\fi}
\def\msunyr{\ifmmode M_{\odot} {\rm yr}^{-1} \else M$_{\odot}$ yr$^{-1}$\fi}
\def\zsun{\ifmmode Z_{\odot} \else Z$_{\odot}$\fi}

\def\lsun{\ifmmode L_{\odot} \else L$_{\odot}$\fi}

\def\mup{\ifmmode M_{\rm up} \else M$_{\rm up}$\fi}
\def\mlow{\ifmmode M_{\rm low} \else M$_{\rm low}$\fi}
\def\aap{A\&A}

\def\apjl{ApJL}
\def\apjs{ApJS}

\newcommand{\oh}{\ifmmode 12 + \log({\rm O/H}) \else$12 + \log({\rm
O/H})$\fi}

\newcommand{\oii}{[O~{\sc ii}]}
\newcommand{\oiii}{[O~{\sc iii}]}
\def\Sii{[S~{\sc ii}] $\lambda\lambda$6717,6731}

\def\Oii{[O~{\sc ii}] $\lambda$3727}
\def\Oiii{[O~{\sc iii}] $\lambda\lambda$4959,5007}
\def\oiiil{[O~{\sc iii}]$\lambda 5007$}
\def\flyf{\ifmmode f_{\rm Lyf} \else $f_{\rm Lyf}$\fi}
\def\pz{\ifmmode P(z) \else $P(z)$\fi}
\def\ki2{\ifmmode \chi^2 \else $\chi^2$\fi}
\def\zphot{\ifmmode z_{\rm phot} \else $z_{\rm phot}$\fi}

\newcommand{\xphot}{\ifmmode x_\gamma \else $v_\gamma$\fi}
\newcommand{\xobs}{\ifmmode x_{\rm obs} \else $x_{\rm obs}$\fi}
\newcommand{\xcmf}{\ifmmode x_{\rm CMF} \else $x_{\rm CMF}$\fi}
\newcommand{\vexp}{\ifmmode V_{\rm exp} \else $V_{\rm exp}$\fi}
\newcommand{\vmax}{\ifmmode V_{\rm max} \else $V_{\rm max}$\fi}
\newcommand{\nh}{\ifmmode N_{\rm HI} \else $N_{\rm HI}$\fi}
\newcommand{\dv}{\ifmmode \Delta v({\rm em-abs}) \else $\Delta v({\rm em}-{\rm abs})$\fi}

\def\fesc{\ifmmode f_{\rm esc} \else $f_{\rm esc}$\fi}
\def\fescrel{\ifmmode f_{\rm esc,rel} \else $f_{\rm esc,rel}$\fi}

\def\frellya{\ifmmode f^{\rm rel}_{\rm{Ly}\alpha} \else $f^{\rm rel}_{\rm{Ly}\alpha}$\fi}

\newcommand{\mstar}{\ifmmode M_\star \else $M_\star$\fi}
\newcommand{\muv}{\ifmmode M_{1500} \else $M_{1500}$\fi}
\newcommand{\auv}{\ifmmode A_{\rm UV} \else $A_{\rm UV}$\fi}
\newcommand{\luv}{\ifmmode L_{\rm UV} \else $L_{\rm UV}$\fi}
\newcommand{\lir}{\ifmmode L_{\rm IR} \else $L_{\rm IR}$\fi}
\newcommand{\lbol}{\ifmmode L_{\rm bol} \else $L_{\rm bol}$\fi}
\newcommand{\liruv}{\ifmmode L_{\rm IR+UV} \else $L_{\rm IR+UV}$\fi}
\newcommand{\liroveruv}{\ifmmode L_{\rm IR}/L_{\rm UV} \else $L_{\rm IR}/L_{\rm UV}$\fi}
\newcommand{\nlyc}{\ifmmode N_{\rm Lyc} \else $N_{\rm Lyc} $\fi}
\newcommand{\rholyc}{\ifmmode \rho_{\rm Lyc} \else $\rho_{\rm Lyc} $\fi}
\newcommand{\chion}{\ifmmode \xi_{\rm ion} \else $\xi_{\rm ion}$\fi}
\newcommand{\chioncorr}{\ifmmode \xi_{\rm ion}^0 \else $\xi_{\rm ion}^0$\fi}

\newcommand{\Civuv}{C~{\sc iv} $\lambda$1550}
\newcommand{\Civ}{C~{\sc iv}}
\newcommand{\Ciii}{C~{\sc iii}]}
\newcommand{\Ciiiuv}{C~{\sc iii}] $\lambda$1909}
\newcommand{\Oiiiuv}{O~{\sc iii}] $\lambda$1666}
\newcommand{\Siiiiuv}{Si~{\sc iii}] $\lambda$1883,1892}

\begin{document}

\title{Strong Lyman continuum emitting galaxies show intense \Civuv\  emission}
\subtitle{}
\author{D. Schaerer\inst{1,2}, 
Y. I. Izotov$^{3}$,
G. Worseck$^{4}$, 
D. Berg$^{5}$, 
J. Chisholm$^{5}$, 
A. Jaskot$^{6}$,
K. Nakajima$^{7}$, 
S. Ravindranath$^{8}$, 
T.X. Thuan$^{9}$,
A. Verhamme$^{1}$
}
%  \offprints{}
  \institute{Observatoire de Gen\`eve, Universit\'e de Gen\`eve, Chemin Pegasi 51, 1290 Versoix, Switzerland
         \and
CNRS, IRAP, 14 Avenue E. Belin, 31400 Toulouse, France
        \and
Bogolyubov Institute for Theoretical Physics,
National Academy of Sciences of Ukraine, 14-b Metrolohichna str., Kyiv,
03143, Ukraine
%Main Astronomical Observatory, Ukrainian National Academy of Sciences,
%27 Zabolotnoho str., Kyiv 03680, Ukraine
        \and
Institut f\"ur Physik und Astronomie, Universit\"at Potsdam, Karl-Liebknecht-Str. 24/25, D-14476 Potsdam, Germany
        \and
Department of Astronomy, University of Texas at Austin, Austin, TX 78712, USA
        \and
Department of Astronomy, Williams College, Williamstown, MA 01267, USA
        \and
National Astronomical Observatory of Japan, 2-21-1 Osawa, Mitaka,
Tokyo 181-8588, Japan
        \and
Space Telescope Science Institute, 3700 San Martin Drive Baltimore, MD 21218, USA
        \and
Astronomy Department, University of Virginia, P.O. Box 400325, 
Charlottesville, VA 22904-4325, USA
         }

\authorrunning{D.\ Schaerer et al.}
\titlerunning{Strong \Civuv\ emission from Lyman continuum emitters}

%\date{Received date; accepted date}
\date{Accepted for publication in A\&A Letters}

%\abstract{CONTEXT}
%{AIMS
%}
%{METHODS
%}
%{RESULTS
%}
%{CONCLUSIONS
%}
%% 5 {} token are mandatory

\abstract{Using the Space Telescope Imaging Spectrograph, we have obtained ultraviolet (UV) spectra from $\sim 1200$ to 2000 \AA\ 
of known Lyman continuum (LyC) emitting galaxies at low redshift ($z \sim 0.3-0.4$) with varying absolute LyC escape fractions ($\fesc \sim 0.01 - 0.72$). 
Our observations include in particular the galaxy J1243+4646, which has the highest known LyC escape fraction at low redshift.
While all galaxies are known Lyman alpha emitters, we consistently detect an inventory of additional emission lines, including \Civuv, \Heiiuv, \Oiiiuv, and \Ciiiuv,
whose origin is presumably essentially nebular.
\Civuv\ emission is detected above 4 $\sigma$ in six out of eight galaxies, with equivalent widths  of EW$($\Civ$)=12-15$ \AA\
for two galaxies, which exceeds the previously reported maximum emission in low-$z$ star-forming galaxies.
We detect \Civuv\ emission in all LyC emitters with escape fractions $\fesc > 0.1$ and find a tentative increase in the flux ratio
\Civuv/\Ciiiuv\ with \fesc. 
Based on the data, we propose a new criterion to select and classify strong leakers (galaxies with $\fesc > 0.1$):  \Civuv/\Ciiiuv\ $\protect\ga 0.75$.
Finally, we also find \Heiiuv\ emission in all the strong leakers with equivalent widths from 3 to 8 \AA\ rest frame.\ These are among the highest values
observed in star-forming galaxies and are primarily due to a high rate of ionizing photon production. 
The nebular \Heiiuv\ emission of the strong LyC emitters does not require harder ionizing spectra at $>54$ eV compared to those
of typical star-forming galaxies at similarly low metallicity.}


 \maketitle

%%%%%%%%%%%%%%%%%%%%%%%%%%%%%%%%%%%%%%%%%%%%%%%%%%%%%%%%%%%%%%%%%%%%%%%%%%%%%%%%%s
\section{Introduction}
\label{s_intro}

In recent years, {\em Hubble Space Telescope} ({\em HST}) observations of star-forming galaxies at low redshift ($z \sim 0.3-0.4$) with the Cosmic Origin Spectrograph (COS)
have measured Lyman continuum (LyC) and the non-ionizing ultraviolet (UV) radiation for nearly 90 galaxies in total, as reported by various studies
\citep[see][]{Leitherer2016Direct-Detectio,Izotov2016Eight-per-cent-,Izotov2016Detection-of-hi,Izotov2018Low-redshift-Ly,Wang2019A-New-Technique,Izotov2021Lyman-continuum} and the Low-$z$ Lyman Continuum Survey \citep[LzLCS;][]{Wang2021The-Low-redshif,Flury2022a,Flury2022b,Saldana2022}.

These data have launched us into a new era, allowing us to study in detail LyC emitting galaxies, their interstellar medium (ISM) properties and stellar populations, 
the conditions that favor the escape of ionizing photons from galaxies, and more \citep[see above references and][]{Schaerer2016The-ionizing-ph,Gazagnes2018Neutral-gas-pro,Chisholm2018Accurately-pred}.
In addition, these observations have served to empirically establish  
``indirect indicators'' of LyC escape, which can potentially be used at all redshifts, including for galaxies in the epoch of reionization, where direct
observations of the LyC are not possible.  Among these indirect indicators are properties of the resolved Lyman alpha (\lya) line profile (the peak separation), 
weak UV absorption lines, Mg~{{\sc ii} emission, a high ratio of \Oiii/\Oii, and a deficit of the \Sii\ line 
\citep[see][]{Jaskot2013The-Origin-and-,Verhamme2017Lyman-alpha-spe,Gazagnes2018Neutral-gas-pro,Henry2018A-Close-Relatio,Chisholm2020Optically-thin-,Ramambason2020Reconciling-esc,Wang2021The-Low-redshif}.

For the exploration of the early Universe and to track the sources of cosmic reionization, the rest-UV spectral range remains fundamental
as it is accessible with the largest ground-based telescopes (and soon with 30 m class telescopes) and with the recently launched {\em James Webb Space Telescope}.
Recent efforts have been undertaken to obtain reference spectra of low-redshift star-forming galaxies covering the range of $\sim 1200-2000$ \AA\
\citep[e.g.,][]{Rigby2015C-III-Emission-,Berg2016carbon-and-Oxyg,Senchyna2017Ultraviolet-spe,Berg2019The-Chemical-Ev}.
These include \lya, \Civuv,  \Heiiuv, O~{\sc iii}] $\lambda\lambda$1660,1666, C~{\sc iii}] $\lambda\lambda$1907,1909, and other emission lines,
which provide important diagnostics of their ISM and radiation field \citep[see, e.g.,][]{Feltre2016Nuclear-activit,Gutkin2016Modelling-the-n,Nakajima2018The-VIMOS-Ultra}.
However, to study sources of reionization and to relate the UV line properties to LyC escape, it is mandatory to observe the same galaxies both in the LyC
and in the non-ionizing UV out to $\sim 2000$ \AA. So far, very few such observations have been obtained, and the first spectrum of a low-$z$ LyC emitter with a
high escape fraction of LyC photons ($\fesc \sim 43$ \%) has been obtained only recently \citep{Schaerer2018Intense-C-III-l}.

Here, we present the first results of an {\em HST} program to observe the UV emission lines of known LyC emitters, sampling a range
of LyC escape fractions from very low ($\fesc = 1.4$ \%) to the highest escape currently known ($\fesc = 72$\%), which was observed in the compact star-forming 
galaxy \object{J1243+4646} by \cite{Izotov2018Low-redshift-Ly}.
Our observations show strong emission lines, in particular strong \Civuv\ emission in the strongest LyC emitters. This demonstrates for the first time
a connection between LyC escape and nebular \Civ\ emission, from which we propose a new empirical criterion to select galaxies with strong LyC escape.
Furthermore, our results shed new light on the recently detected \Civ\ emitters at high redshift
\citep[see][]{Stark2015Spectroscopic-d,Mainali2017Evidence-for-a-,Schmidt2017The-Grism-Lens-,Tang2021Rest-frame-UV-s,Vanzella2021The-MUSE-Deep-L,Richard2021An-atlas-of-MUS}, 
suggesting that some of them could be strong LyC emitters.

%%%%%%%%%%%%%%%%%%%%%%%%%%%%%%%%%%%%%%%%%%%%%%%%%%%%%%%%%%%%%%%%%%%%%%%%%%%%%%%%%
\section{UV spectra of compact star-forming galaxies at $z \sim 0.3-0.4$  with varying LyC escape fractions}
\label{s_obs}

\subsection{HST observations}
Eight out of eleven compact star-forming galaxies with LyC measurements from our 2016--2018 campaign 
\citep{Izotov2016Eight-per-cent-,Izotov2016Detection-of-hi,Izotov2018J11542443:-a-lo,Izotov2018Low-redshift-Ly}
have been observed in Cycle 27 (GO 15941; PI Schaerer) and earlier \citep{Schaerer2018Intense-C-III-l}\footnote{The galaxies are \object{J1243+4646},  \object{J1154+2443},  \object{J1152+3400},  \object{J1442-0209},  \object{J0925+1403},  \object{J1011+1947},  \object{J0901+2119}, and  \object{J1248+4259}.}.
The sources span a broad range of LyC escape fractions ($\fesc \sim 1.4-72$ \%), metallicities
in the range $\oh = 7.64 - 8.16$ with a median of 7.92, and \oiiil/\Oii\  from 5 to 27.1.

The observations were taken with the Space Telescope Imaging Spectrograph (STIS) NUV-MAMA using
the grating G230L with the central wavelength 2376 \AA\ and the slit 52\arcsec $\times$ 0\farcs 5,
resulting in a spectral resolution of $R \sim 700$ for the targeted compact galaxies.
The data were reduced with the CALSTIS v3.4.2 pipeline and including data from additional observations of  \object{J1154+2443} that were not available to \cite{Schaerer2018Intense-C-III-l}, which improved the signal-to-noise ratio (S/N) to $\simeq 7$ per pixel for this galaxy.
% using the default 11-pixel source extraction box, but larger 100-pixel background extraction windows placed 70 pixels away from the trace. 
Wavelength shifts between sub-exposures were insignificant, which allowed for the co-addition of gross and background counts per pixel, approximately preserving Poisson statistics. 
%The background-subtracted counts were converted to flux, and one-sigma Poisson flux uncertainties were calculated using the approach by http://adsabs.net/abs/1998PhRvD..57.3873F. 
In the wavelength range of interest, the continuum S/N (accounting for Poisson flux uncertainties) ranges between 3 and 8 per $\simeq 1.55$\,\AA\ pixel, with six out of eight spectra reaching S/N$>5$.

For the remainder of this Letter, we define ``strong'' leakers as galaxies with a LyC escape fraction
above 10 \% (i.e., $\fesc > 0.1$) to distinguish such sources, which could significantly contribute to 
cosmic reionization, from those with a low or negligible escape of ionizing photons. 
In our current sample, three out of eight galaxies are classified as  strong leakers (see Fig.~\ref{fig_c4c3_oh}).
%\LEt{ Since you don't label the panels as "1" and "2", I\ suggest changing this to "the right/left panel of Fig.\ 2" here and throughout.}. % Fig 3

\setcounter{figure}{0}
\begin{figure}[htb]
{\centering
\includegraphics[width=9.9cm]{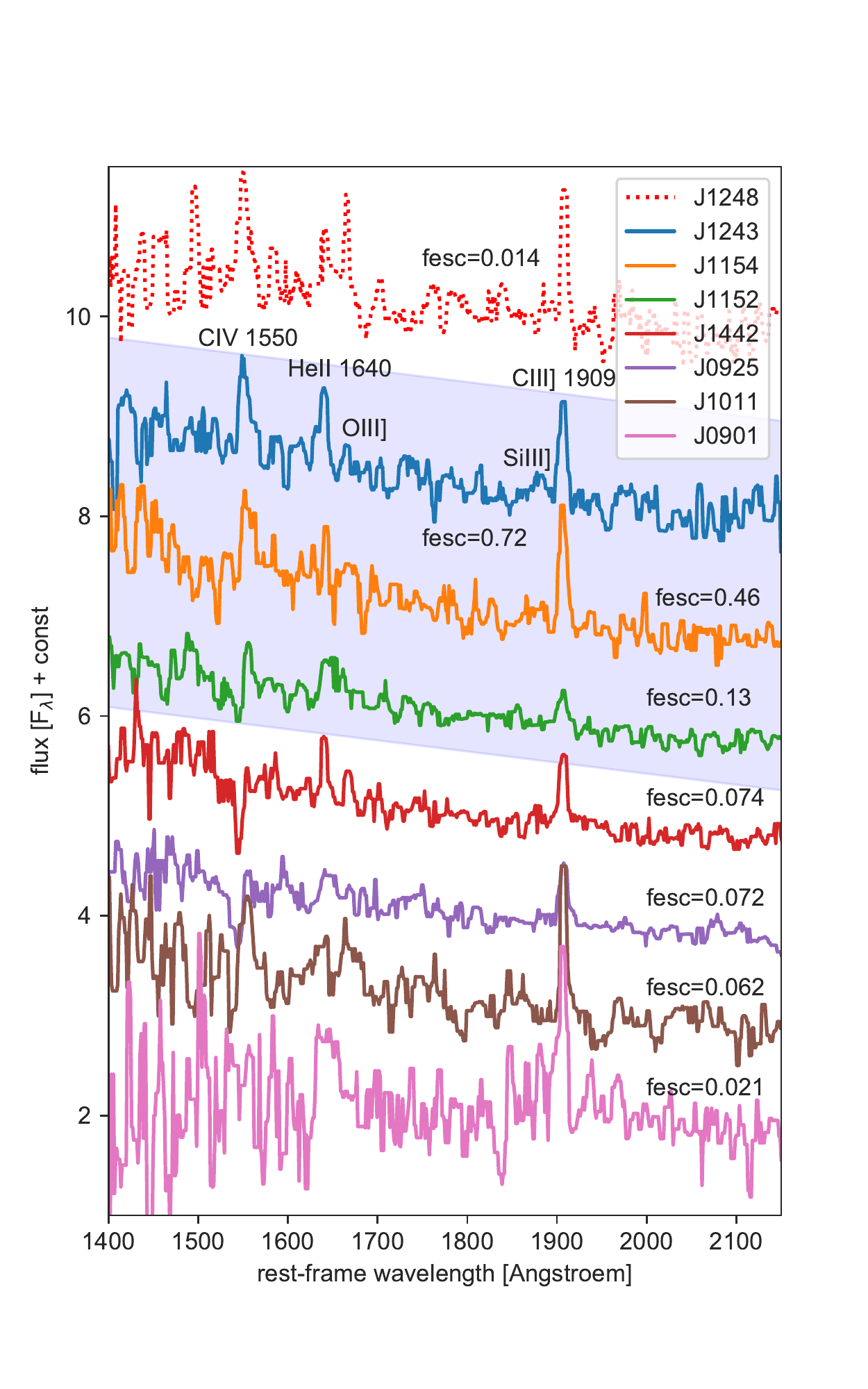}
\caption{{STIS rest-frame spectra of eight $z \sim 0.3-0.4$ LyC emitters. They are ordered by decreasing \fesc\ from top to bottom,
except for J1248+4259, which is shown at the top with a dotted line (see text for details). The colored area shows the spectra of the three strong LyC leakers (\fesc $>0.1$).
The \Civuv, \Ciiiuv, and \Heiiuv\ lines are clearly detected in the strongest LyC leakers ( \object{J1243+4646},  \object{J1154+2443}, and  \object{J1152+3400},
with escape fractions between $\fesc=0.13$ and 0.72), which are characterized by a high C~{\sc iv}/\Ciii\ ratio.
All the spectra show \Ciiiuv\ emission (with EWs of $\sim 5-20$ \AA\ rest frame) and other weaker lines.}
}}
\label{fig_spectra}
\end{figure}

\subsection{Main UV emission line features}

Examples of the observed  STIS spectra are shown in Fig.\ \ref{fig_spectra},
ordered -- from top to bottom -- by decreasing LyC escape fraction.
The top two sources shown there have very high escape fractions.
In the spectral range covered by the observations (rest-frame wavelengths of $\sim 1200 - 2000$ \AA),
the main detected emission lines are \lya\ (not illustrated here), \Civuv, \Heiiuv, \Oiiiuv, and \Ciiiuv.
We note that with the given resolution, the \Civuv\ and \Ciiiuv\ doublets are not resolved.
Hints of blended \Siiiiuv\ are also visible in some of the spectra (see Fig.~\ref{fig_spectra}).

We now mainly focus on the carbon and helium  lines, and leave a more exhaustive report and analysis for subsequent publications.
We note that the low resolution of our spectra does not allow us to exclude some contribution from stellar emission (from O, B, and Wolf-Rayet stars)
to the \Civuv\ and \Heiiuv\  emission lines. However, we think that the emission is predominantly nebular in both cases, at least for sources 
with high equivalent widths (EWs) (EW$($\Civ$) \ga 3$ \AA), since absorption from stellar P-Cygni \Civ\ profiles is not detected in high EW sources and since no Wolf-Rayet  signatures are seen
 in the available deep optical spectra of LyC emitters \citep{Guseva2020Properties-of-f}.

After \lya, the lines with the highest EWs are \Ciiiuv\ and  \Civuv, with EW$($\Ciii$)=4-20$ \AA\ and 
EW$($\Civ$)=3-15$ \AA. The \Ciiiuv\ line is detected in all the sources, and \Civuv\ in five out of eight galaxies above 4 $\sigma$.

In Fig.~\ref{fig_ewc3_oh} 
we plot the \Ciii\ and \Civ\ EWs of our targets as a function of metallicity (using O/H as a proxy) along with 
measurements from other low-redshift galaxies for comparison.
The \Ciii\ EWs of the LyC emitters do not occupy a particularly unique domain in this figure, and EW$($\Ciii$)$ does not show a dependence on the LyC escape fraction.
The ``envelope'' of the observed distribution of the \Ciii\ EWs show a metallicity dependence 
that is fairly well reproduced by photoionization models using the spectral energy distributions of young stellar populations, as shown 
in several studies \citep{Jaskot2016Photoionization,Nakajima2018The-VIMOS-Ultra,Ravindranath2020The-Semiforbidd}.

Interestingly, the \Civ\ EWs of the strong leakers are among the highest observed so far.
Furthermore,  \object{J1248+4259} stands out as the low-$z$ galaxy ($z \la 0.4$) with the highest \Civuv\ EW (EW$($\Civ$)=15.14 \pm 2.14$ \AA);
however, it shows a low escape fraction ($\fesc = 0.014 \pm 0.004$).
Again, the observed EWs are also in fair agreement with the predictions from photoionization models \citep{Nakajima2018The-VIMOS-Ultra,Ravindranath2020The-Semiforbidd}.

\Heiiuv\ emission is detected at $\sim 4-6 \, \sigma$ in all strong leakers, showing EWs of EW$($\heii$) = 3.7 - 8.0$ \AA,
which are among the highest values observed in star-forming galaxies.
The \heii\ line is also detected  above $3 \sigma$ in two other sources (i.e.,\ in five of the eight spectra).
We comment on the strength of the \Heiiuv\ emission below.

% % % % % % % % % % % % % % % % % % % % % % %
\begin{figure*}[htb]
{\centering
\includegraphics[width=9cm]{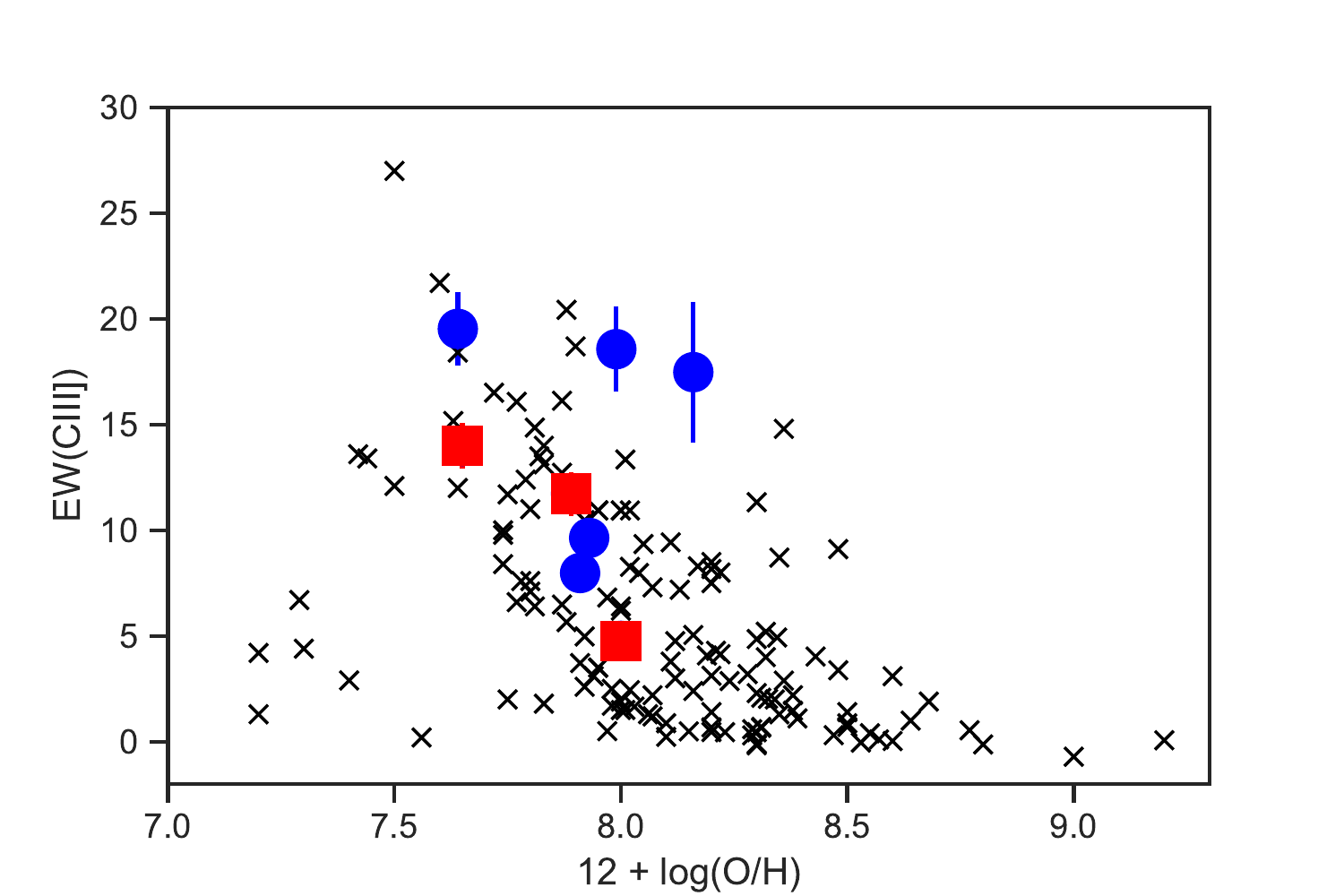}
\includegraphics[width=9cm]{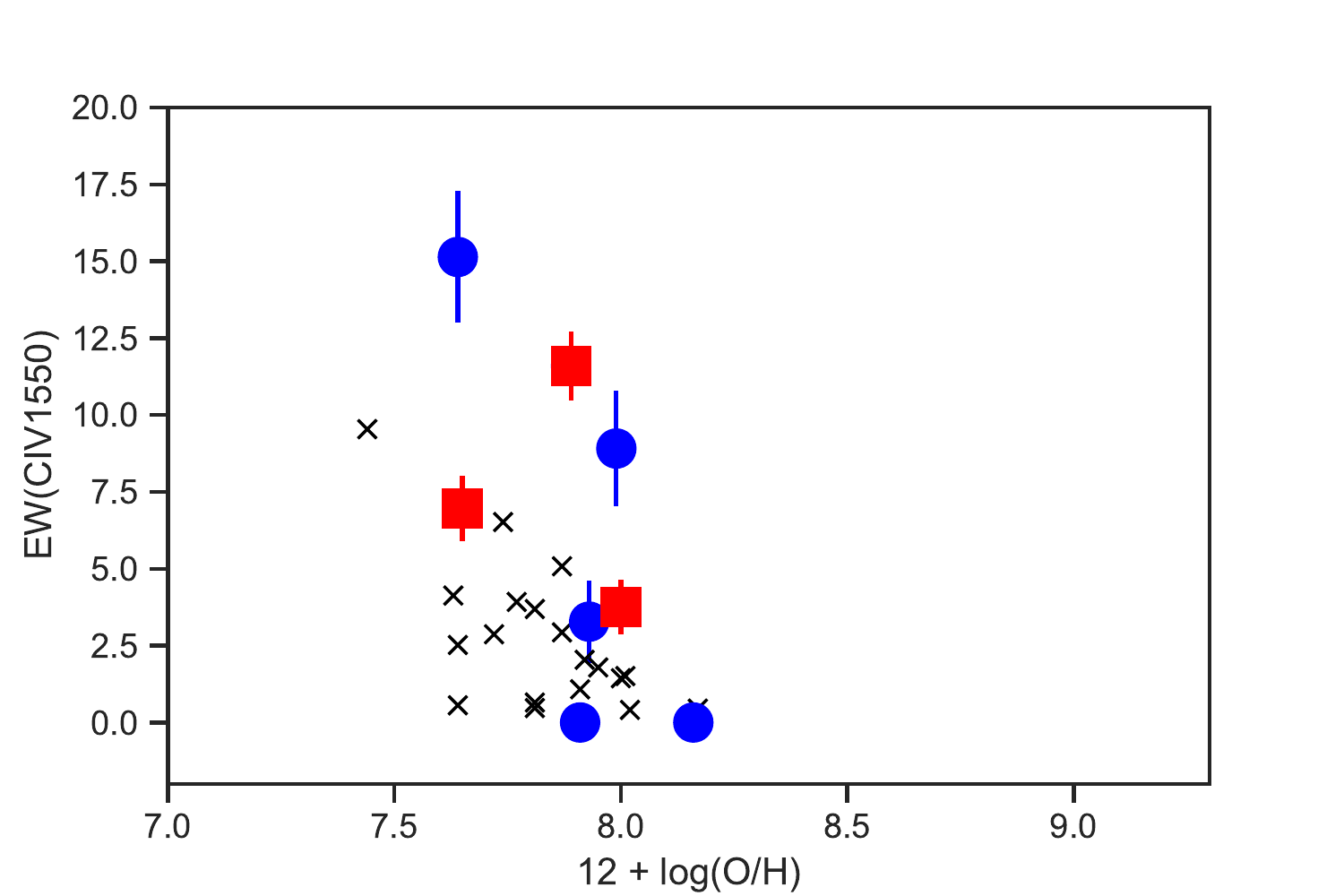}
\caption{Rest-frame \Ciiiuv\ (left panel) and \Civuv\ (right panel) EWs as a function of metallicity O/H for the LyC emitters (large red squares: strong leakers, $\fesc > 0.1$; blue circles: other leakers) and low-$z$ comparison samples from the literature taken from the compilation of \cite{Schmidt2021Recovery-and-an}.
All measurements are for galaxies with direct metallicity measurements.
The typical relative errors for our sources are 10--20 \% for all EW$($\Ciii$)$ and for EW$($\Civ$) \protect\ga 5$ \AA, and larger otherwise (shown).}
}
\label{fig_ewc3_oh}
\end{figure*}

\begin{figure*}[tb]
{\centering
\includegraphics[width=9cm]{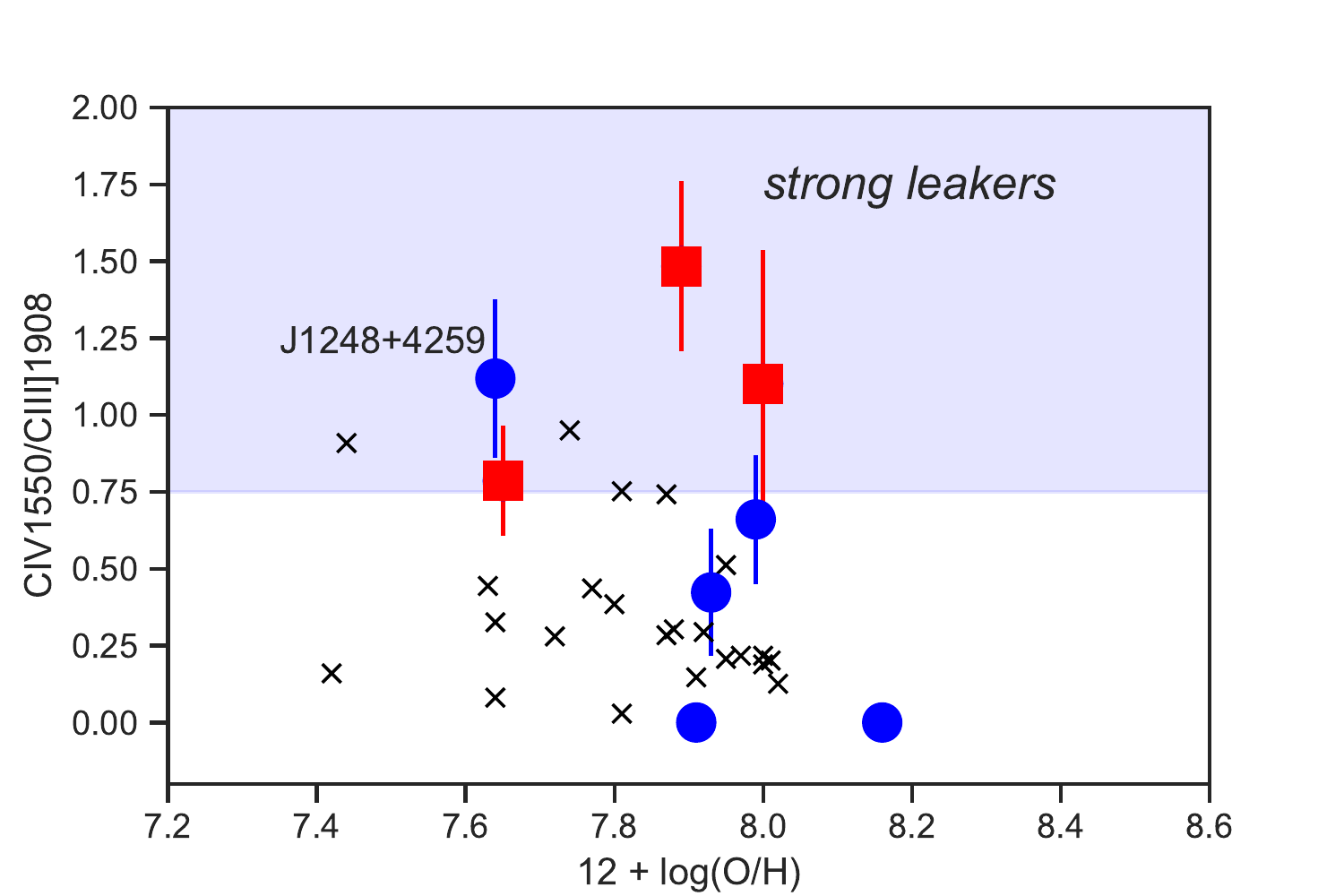}
\includegraphics[width=9cm]{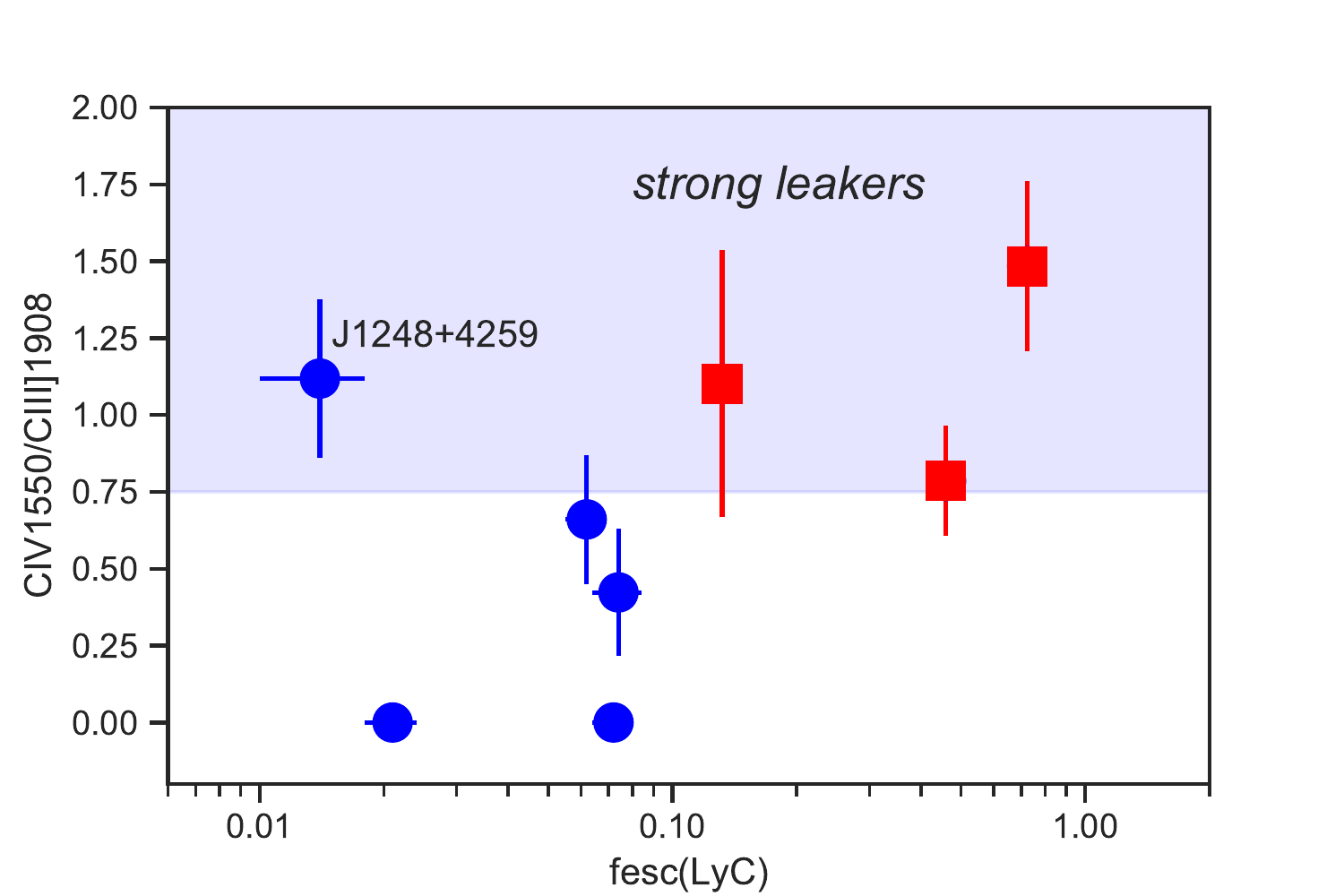}
\caption{\Civuv/\Ciiiuv\ line ratio (C43) as a function of metallicity (O/H, left panel) and the LyC escape fraction (right). The symbols have the same meaning as in Fig.~\ref{fig_ewc3_oh}. 
The colored area shows the region of C43$\protect\ga 0.75$, where we propose that strong LyC leakers, i.e.,\ galaxies with $\fesc >0.1$, are found empirically.
We note that for the sake of simplicity for comparisons with high-$z$ studies the line ratios of the leakers are not corrected for internal reddening and could thus be slightly higher, 
although the corrections should be small.}}
\label{fig_c4c3_oh}
\end{figure*}
% % % % % % % % % % % % % % % % % % % % % % %

% % % % % % % % % % % % % % % % % % % % % % %
\subsection{Strong LyC leakers show strong  \Civuv\ emission and a high C~{\sc iv}/\Ciii\ ratio}

In addition to exhibiting strong \Civ\ and \Heiiuv\ emission lines,
the main distinguishing feature of the strong leakers is the high ratio, C43 $= I($\Civuv$)/I($\Ciiiuv$) \ga 0.75,$ of the carbon line intensities,
which is fairly exceptional, as shown in Fig.~\ref{fig_c4c3_oh}
%\ref{fig_c4c3_oh} 
(left panel), in comparison with other low-$z$ galaxies where both carbon
lines have been detected.
And interestingly, except for the galaxy  \object{J1248+4259} discussed below, the carbon line ratio increases with the LyC escape fraction, as shown in the right panel.
Although the sample of galaxies for which both observations of the LyC and the non-ionizing UV spectrum, including \Ciii\ and \Civ, are covered
is admittedly small, we suggest that, based on the available data, star-forming galaxies with $I($\Civuv$)/I($\Ciiiuv$) \ga 0.75$
have escape fractions $\fesc > 0.1$ and hence show ``significant'' amounts of LyC emission.

If generally applicable, our postulate implies that the galaxy  \object{J1248+4259} reported by \cite{Izotov2018Low-redshift-Ly} should be a strong LyC emitter; that is, it should\ have a true LyC escape fraction of at least $\sim 5$ times that measured from the COS spectrum, which could be possible\ if, for example,
the emission of LyC photons is not isotropic.
There are indeed other indications that  \object{J1248+4259 }could be a strong leaker: First, it shows a very strong \lya\ emission with an EW
of EW$(\lya)=256 \pm 5.2$ \AA, comparable to strong leakers. Second, the \lya\ line profile is clearly double-peaked, with a fairly low separation of the two peaks
($v_{\rm sep} = 283.8 \pm 15.9$ \kms), which would yield $\fesc = 0.13$ if the mean relation between \fesc\ and $v_{\rm sep}$ from  \cite{Izotov2018Low-redshift-Ly} 
was applied. Finally,  \object{J1248+4259} is also somewhat of an ``outlier'' in the scattered relation between O32$=$\oiii/\oii\ and \fesc, showing a lower-than-average escape fraction for
its high O32=11.8 \citep[see][]{Izotov2021Lyman-continuum}.

The other low-$z$ galaxies that show $I($\Civuv$)/I($\Ciiiuv$) \ga 0.75$ in Fig.~\ref{fig_c4c3_oh} 
are  \object{J084236} and  \object{J104457}. They are low-mass low-metallicity galaxies at $z \sim 0.01$ that both show
very strong \Civ\ emission with EW$($\Civ$) \sim 6-10$ \AA\  \citep{Berg2019The-Chemical-Ev,Berg2019Intense-C-IV-an}.
Based on high-resolution COS spectra, which show strong nebular \Civ\ emission in the two doublet lines and indications for
minor radiation transfer effects in these lines,  \cite{Berg2019Intense-C-IV-an} have suggested that  \object{J104457} is optically thin to ionizing radiation, at least at energies
above 47.9 eV, the ionization potential of C$^{3+}$ \citep[see also][]{Senchyna2021Direct-Constrai}.
It also shares other properties of strong LyC leakers, such as a very high O32$=16.2$.
Although no direct observations of the LyC are possible at these low redshifts, the available data for  \object{J104457} appear fully consistent with 
our postulate that this galaxy is a strong LyC leaker. We suggest that  \object{J084236} may also be a strong LyC leaker.

Intrinsically,  the C43 ratio traces the ionization structure of the nebula, similarly to O32, which has already been suggested as a potential
indicator of LyC escape \citep{Jaskot2013The-Origin-and-,Nakajima2014Ionization-stat}. However, since \Civuv\ is a resonance line, it is a priori affected 
by radiation transfer effects; this is in contrast to the nebular oxygen lines, which should be optically thin. Therefore, the \Civuv\ line and the C43 ratio could be a  more sensitive tracer of LyC escape than O32.
To first order, both C43 and O32 trace the ionization parameter, $U$, and are independent of metallicity. 
Furthermore, an examination of photoionization models shows that both C43 and O32 vary with the LyC escape fraction.
For moderate to high $U$, the predicted variations in C43 appear stronger than in O32, again providing some theoretical support for our
empirically based postulate. A detailed comparison with models will be presented in future work.

%%%%%%%%%%%%%%%%%%%%%%%%%%%%%%%%%%%%%%%%%%%%%%%%%%%%%%%%%%%%%%%%%%%%%%
\section{Discussion and implications}
\label{s_discuss}

\subsection{Comparison with high-$z$ LyC emitters and candidates}

We first compared our results with confirmed high-$z$ LyC emitters, although
relatively few high-$z$ galaxies with established or potentially strong LyC escape (absolute escape fractions $\fesc \ga 0.1$) are known.
Well-studied sources with the highest LyC escape fractions are {\em  \object{Ion2}}, the {\em  \object{Sunburst}} arc, and probably {\em  \object{Ion3}}, with 
relative escape fractions $\fescrel \ga 0.5$, according to \cite{Vanzella2020Ionizing-the-in}\footnote{We here refer to relative escape fractions since
absolute values cannot reliably be determined for individual sources at high $z$ \citep[e.g.,][]{Steidel2018The-Keck-Lyman-}.}.
% STRONG LEAKERS:
VLT/XShooter spectroscopy of {\em Ion2} at $z=3.2$ indeed shows the presence of nebular emission in both \Civ\ and \Ciii, with a ratio 
$I($\Civuv$)/I($\Ciiiuv$) = 0.61 \pm 0.23$; this is compatible with our findings, 
although the lines are relatively weak \citep[e.g.,\ EW$($\Civ$)=2.6$ \AA\ and EW$($\heii$)=2.8$ \AA ;][]{Vanzella2020Ionizing-the-in}.
For the other objects, the reported data are insufficient to examine \Civ\ and \Ciii.
Similarly, no detailed rest-UV spectra have been published for the $z=3.1$ LyC emitting galaxies of \cite{Fletcher2019The-Lyman-Conti}.
The robust sample of $ z \sim 3$ LyC emitters from \cite{Steidel2018The-Keck-Lyman-} and \cite{Pahl2021An-uncontaminat} provides average properties
to avoid the inherent limitations in determining \fesc\ for individual sources that are due to the stochastic intergalactic medium transmission at high $z$.
Their LyC emitting sample has  an average escape fraction of $\fesc = 0.06 \pm 0.01$, which is below our definition of strong emitters, and the stacked UV spectrum
illustrated in \cite{Steidel2018The-Keck-Lyman-} does not show strong \Civ\ emission, although no details  on carbon emission lines are reported in their study.

Another comparison is possible with \lya \ emitters at $z \sim 2$, where the LyC escape fraction was recently estimated using indirect empirical 
spectral indicators by \cite{Naidu2021The-Synchrony-o} and \cite{Matthee2021ReSolving-Reion}. Stacking rest-UV spectra of \lya \ emitters with estimated 
$\fesc \ga 0.2$ (high escape) and $\fesc \la 0.05$ (low escape), they find several significant differences  between the two subsamples,
in particular the presence of nebular \Civuv, \Ciiiuv, \Heiiuv, and \Oiiiuv\ emission in the high escape sources, whereas the low escape sources only 
show \Ciii\ and O~{\sc iii}] emission.
Although the \Civ\ emission of their strong leakers (EW$($\Civ$)=2.0 \pm 0.4$ \AA) is weaker than that of our three sources with $\fesc > 0.1,$
which show EW$> 2.5$ \AA,  and despite \Ciiiuv\ being stronger than \Civ, 
%(by a factor of $\sim 3.4$, according to the reported EW ratio), 
their finding of nebular \Civ\ and \Heiiuv\ emission in strong leaker candidates is compatible with our results.

We therefore conclude that both low- and high-redshift observations seem to consistently show (i.e., in about three out of four cases) that the UV spectra of strong leakers (with $\fesc \ga 0.1-0.2$) are characterized by nebular \Civuv\ and \Ciiiuv\ emission and C43 $\ga 0.75$. Although the available data are still relatively scarce, our observations provide the first such quantitative estimate.

%\vspace*{-0.3cm}
\subsection{Implications for high-$z$ CIV emitters}
So far, very few high-redshift galaxies showing nebular \Civuv\ emission have been reported. Since the recent discoveries of several
lensed galaxies at $z \sim 6-7$ with nebular \Civ\ \citep{Stark2015Spectroscopic-d,Mainali2017Evidence-for-a-,Schmidt2017The-Grism-Lens-},
other \Civ\ emitters have been found both in blank fields \citep[e.g.,\ at $z \sim 2.2$;][]{Tang2021Rest-frame-UV-s} {and in lensed galaxies
\citep[see][]{Stark2014Ultraviolet-emi,Vanzella2021The-MUSE-Deep-L,Richard2021An-atlas-of-MUS}. }
The stacked rest-UV spectrum of lensed sources with
median $z=3.2$ and median $\muv = -17.1$ from \cite{Vanzella2021The-MUSE-Deep-L} shows strong nebular \Civuv, \Ciiiuv, and numerous other
emission lines. The 
\Civ\ line in several of these sources is strong, with EWs of up to EW$($\Civ$)=38 \pm 16$ \AA\ in A1703-zd6 
\citep{Stark2015Spectroscopic-d,Schmidt2021Recovery-and-an}
and C43 $>0.8$ in several of them. 
Our results suggest that these sources are strong LyC emitters, which is overall in agreement with the other observational
properties they share with LyC emitters (e.g.,\ strong \lya\ emission). The fact that \Civ\ emission is found in many strongly lensed sources
probably also indicates that the fraction of strong LyC emitters increases toward fainter, lower-mass galaxies, a trend also found 
in the $z \sim 0.3-0.4$ reference samples \citep[see][]{Izotov2021Lyman-continuum,Flury2022b}.

%\vspace*{-0.4cm}
\subsection{Strong emission lines and high C43 ratios explained}
This leads us to the questions of what the strong observed UV emission lines (high EW of nebular lines) tell us and how they can be understood.
Since EWs measure the strength of the emission line with respect to the continuum, the EW
of a UV recombination line such as \Heiiuv\ can be cast in the following simple form,
\begin{equation}
EW(1640)/\AA= \chion^H  \frac{Q_{\rm He+}}{Q_{\rm H}} \frac{c_{1640} \lambda^2}{3\times 10^{18}} , 
\end{equation}
where $\chion^H = Q_{\rm H} / L_{1500}$ is the ionizing photon efficiency per unit intrinsic monochromatic UV luminosity
in the commonly used units of erg$^{-1}$ Hz,  $c_{1640} = 5.67 \times 10^{-12}$ erg relates the recombination line flux 
to the ionizing photon flux \citep[e.g.,][]{schaerer2003}, 
and $\lambda \approx 1640$ \AA .
This shows that the EW is proportional to \chion\ and to the hardness of the ionizing spectrum, expressed here
as $Q_{\rm He+}/Q_{\rm H}$, the ratio of the ionizing photon flux above 54 eV and 13.6 eV, respectively.
From this we can easily see that the observed EWs in our three strongest leakers, EW(1640)$ \approx 3-8$ \AA,\ are 
plausible. Indeed, the EWs can be explained with a relatively high ionizing photon production, $\log(\chion) \approx 25.6-25.8$
erg$^{-1}$ Hz \citep[cf.][]{Schaerer2018Intense-C-III-l}, which then implies $Q_{\rm He+}/Q_{\rm H} \approx 0.01-0.04$. This hardness translates to optical 
line ratios of $I($\Heiiopt$)/I(\hb) \approx 1.74 \, Q_{\rm He+}/Q_{\rm H} \sim 0.016 - 0.07$, similar to the observations of  \object{J1154+2443} and  \object{J1248+4259},
which are $I(4686)/I(\hb) \sim 0.02-0.03$ \citep{Guseva2020Properties-of-f}, and comparable to the line intensities typically observed in galaxies at comparable 
metallicity \citep[at $\oh \sim 8.0$; see, e.g.,][]{Schaerer2019X-ray-binaries-}.
This simple estimate shows that the high \Heiiuv\ EWs reached in these leakers do not require exceptional conditions and/or 
exceptionally hard ionizing spectra compared to other galaxies (weak or non-leakers) at similarly low metallicity. 
The same conclusion also applies\ to, for example, the strong leaker candidates found by  \cite{Naidu2021The-Synchrony-o} among 
their \lya\ emitter sample. 
However, the observed spectra are clearly harder at $> 54$ eV than predicted by normal stellar population models, and
 the source of these He$^+$-ionizing photons 
remains to be elucidated \citep[see, e.g.,][]{Stasinska2015Excitation-prop}. 

To predict the strengths of the metallic lines of \Civ\ and \Ciii, which depend on the ionization parameter, spectral energy distribution, 
nebular abundances, and other factors, detailed photoionization models are required.
The models of \cite{Nakajima2018The-VIMOS-Ultra}, for example, show maximum values of EW$($\Civ$)\sim 8-10$ \AA\ and
EW$($\Ciii$)\sim 10-18$ \AA\ at metallicities $\oh \sim 7.7-8$, of the same order as our observed EWs.
We also examined density-bounded models at appropriate metallicities, using the {\em BOND} set of CLOUDY models from the 3 Million Models Database
of \cite{Morisset2015A-virtual-obser}. The models indeed show an expected increase in the \Civ/\Ciii\ ratio with increasing 
LyC escape fraction, and several models show \Civ/\Ciii\ ratios comparable to those of our observations.
From these simple comparisons, we conclude that both the observed EWs and UV line ratios of the strong low-$z$ leakers
seem ``reasonable''  and that we probably do not need peculiar or extreme ionizing spectra to reproduce their emission line
spectra. 
Tailored photoionization models  to examine the behavior of the major UV and optical emission lines of the LyC emitters
and comparison sources will be presented in a subsequent publication.

%%%%%%%%%%%%%%%%%%%%%%%%%%%%%%%%%%%%%%%%%%%%%%%%%%%%%%%%%%%%%%%%%%%%%%
\section{Conclusion}
\label{s_conclude}

With the STIS spectrograph on board {\em HST,} we have obtained new UV spectra from $\sim 1200-2000$ \AA, the\ rest-frame spectra of eight $z \sim 0.3-0.4$ LyC emitters 
from \cite{Izotov2016Eight-per-cent-,Izotov2016Detection-of-hi,Izotov2018J11542443:-a-lo,Izotov2018Low-redshift-Ly}, which cover a large range of LyC escape fractions.
We detect multiple emission lines, including \lya, \Civuv, \Heiiuv, \Oiiiuv, and \Ciiiuv. The \lya\ and \Ciii\ lines are detected in all the sources.
Our main results can be summarized as follows:
\begin{itemize}
\item We report the detection above 4 $\sigma$ of  \Civuv\ emission in six out of eight galaxies,  with the EWs in two galaxies ( \object{J1243+4646} and  \object{J1248+4259}),  EW$($\Civ$)=12-15$  \AA,\
exceeding the previously reported maximum in low-$z$ galaxies \citep[ \object{J104457}; ][]{Berg2019Intense-C-IV-an}.
\item Strikingly, \Civuv\ emission is detected in all LyC emitters with escape fractions $\fesc > 0.1$, and the flux ratio of
\Civuv/\Ciiiuv\ appears to increase with \fesc. 
\item Based on the available data, we suggest that strong leakers, galaxies with $\fesc > 0.1$, are characterized by \Civuv/\Ciiiuv\ $\ga 0.75$,
adding another indirect indicator of LyC escape to those already established.
\item All strong leakers also show strong \Heiiuv\ emission with EW$($\heii$) = 3.7 - 8.0$ \AA, which are among the highest values observed in star-forming galaxies.
\item A simple estimate shows that the high EW of the \Heiiuv\ line is primarily due to a high ionizing photon production, \chion,
and that it does not require unusually hard ionizing spectra, compared to normal galaxies at similar metallicity that frequently show
optical \Heiiopt\ emission lines.
\end{itemize}

In short, our observations provide  an important new reference for understanding the UV spectra of LyC emitting galaxies and thus also
analogs of the sources of cosmic reionization.
The spectra of the strong low-$z$ leakers share many similarities with those of the \Civuv\ emitters recently discovered at high redshifts
\citep[e.g.,][]{Stark2015Spectroscopic-d,Mainali2017Evidence-for-a-,Schmidt2017The-Grism-Lens-},
and our results suggest that these objects are good candidates for strong LyC escape.
If universally applicable, the empirical criterion of using the carbon line ratio \Civuv/\Ciiiuv\ $\ga 0.75$ to identify strong LyC leakers
could represent an additional important tool for studying the sources of cosmic reionization.

%%%%%%%%%%%%%%%%%%%%%%%%%%%%%%%%%%%%%%%%%%%%%%%%%%%%%%%%%%%%%%%%%%%%%%%%%%%%%%%%%
\begin{acknowledgements}
Y.I. acknowledges support from the National Academy of Sciences of Ukraine by its priority project  ``Fundamental properties of the matter in relativistic collisions of nuclei and in the early Universe''.
D.B., J.C, A.J., and T.X.T. acknowledge support from grant HST-GO-15941.002-A. 
This research is based on observations made with the NASA/ESA Hubble Space Telescope obtained from the Space Telescope Science Institute, which is operated by the Association of Universities for Research in Astronomy, Inc., under NASA contract NAS 5-26555. These observations are associated with programs GO 15941 and GO 15433 (PI Schaerer).

\end{acknowledgements}
%%%%%%%%%%%%%%%%%%%%%%%%%%%%%%%%%%%%%%%%%%%%%%%%%%%%%%%%%%%%%%%%%%%%%%%%%%%%%%%%%
\bibliographystyle{aa}
%\bibliography{references}
\bibliography{merge_misc_highz_literature}

\end{document}